# Harnessing diamond surface features for dense and aligned NV ensembles

Eveline Postelnicu[a,*], Lillian B. Hughes Wyatt[b], Tri Nguyen[a], Simon A. Meynell[c], Christine Jilly[d], Paul Wallace[e], Andrew Barnum[e], Ania Bleszynski Jayich[c], Kunal Mukherjee[a,*]

[a]*Department of Materials Science and Engineering, Stanford University, Stanford, California 94305, USA*

[b]*Materials Department, University of California, Santa Barbara, Santa Barbara, California 93106, USA*

[c]*Department of Physics, University of California, Santa Barbara, Santa Barbara, California 93106, USA*

[d]*Stanford Doerr School of Sustainability - Dean's Office, Stanford, California, 94305, USA*

[e]*nano@stanford, Stanford, California, 94305, USA*

*Email: eveline.postelnicu@gmail.com

*Email: kunalm@stanford.edu

**Abstract:**

Controlling nitrogen doping in diamond is key to advancing nitrogen-vacancy (NV) center devices. We harness the hillock, a typically undesirable surface feature, to incorporate high densities of grown-in, aligned NV-centers on a (001)-oriented substrate. Enhanced cathodoluminescence at hillock sidewalls is correlated via nanoSIMS to up to 1000× greater nitrogen incorporation compared to the planar film. We find that these hillocks are associated with stacking faults and edge-type dislocations, consistent with an origin in surface preparation rather than substrate screw dislocations. Yet, the growth is orderly enough that each of the four hillock sidewalls hosts a distinct NV orientation. A 1.7–2% grown-in NV/substitutional nitrogen (P1) ratio, 4× higher than typical (001)-oriented growth, is measured via NV decoherence analysis. By revealing that spontaneously formed hillocks act as natural laboratories for dense, aligned NV formation, this work motivates systematic investigation of facet-dependent nitrogen incorporation and preferential NV alignment in (001) diamond.

**Manuscript:**

High quality synthesis of diamond thin films via epitaxy is crucial for solid state spin qubits for quantum technologies with applications in sensing, simulation, and networking.[1–5] Understanding nitrogen doping is of particular interest for the nitrogen-vacancy (NV) center, a promising and well-explored qubit system. In widely used (001)-oriented diamond, the NV center offers long coherence times and room-temperature optical addressability.[6–8] Looking ahead, dense, aligned ensembles of NVs can increase sensitivity and harness strong interactions between NVs for entanglement enhancement in sensing and quantum simulation.[9–14] Today, such ensembles are often generated either by annealing heavily nitrogen ion-implanted samples or by annealing CVD-grown, heavily nitrogen-doped films after electron irradiation.[15–18] Both techniques have their downsides. The NV density is limited by nitrogen incorporation for CVD growth on (001) surfaces while ion implantation leads to additional damage that can degrade coherence.[12] Moreover, thermal annealing leads to a random distribution of all four possible NV orientations as vacancies diffuse indiscriminately to substitutional nitrogen atoms.[19] Grown-in NVs on a single (001) surface also typically exhibit all four possible NV orientations and a low substitutional nitrogen-to-NV conversion rate of 0.5%.[11,20] Studies of nitrogen incorporation on non-(001) surfaces, such as {111} and {113}, show both greater incorporation during CVD growth and the possibility of preferentially aligned grown-in NV centers, but these orientations are harder to procure commercially and harder to prepare.[21]

Surface nonuniformities on the (001)-surface such as hillocks and step bunches, long recognized as characteristic features in CVD diamond films, may provide an approach to understand aspects of growth available at other orientations like high density and preferential alignment. We ask how nonuniformities affect nitrogen incorporation and the formation of grown-in NVs and whether these surface features can act as natural laboratories for observing dense, aligned NV centers on (001)-oriented templates. Growth hillocks in CVD diamond epitaxial films originate from grown-in dislocations in the diamond substrates or surface contaminants introduced during growth (e.g., from the plasma).[22] Hillocks are typically considered undesirable for qubits because of uncontrolled surface morphology and variable strain fields associated with these extended defects that can lead to inhomogeneous broadening of spin transitions and reduced $T_2^*$ coherence.[23] In this work, we revisit this assumption and show that hillocks can exhibit strongly enhanced local nitrogen incorporation with highly preferentially oriented grown-in NV-centers in the hillock facets, with coherence only limited by substitutional nitrogen.

We prepared two single-crystalline diamond thin films grown on 001-oriented diamond substrates: a “low miscut” sample with 0.60° polar and -33° from [110] azimuth miscut, and a “high miscut” sample with 1.25° polar miscut towards [110]. The samples are grown by plasma-enhanced chemical vapor deposition (PECVD) in a Seki reactor and intentionally doped with nominally 7–15 nm thick $^{15}$N layer(s) during growth. Following growth, the samples exhibited a variety of common surface features such as hillocks and step bunches. Fig. 1 shows growth schematics (Fig. 1a,e) for the two samples as well as scanning electron microscope (SEM) secondary electron (Fig. 1b,f) and panchromatic cathodoluminescence (CL) (Fig. 1c,g) images of the sample surfaces. Atomic force microscopy (AFM) topography maps are also shown for hillocks analyzed further throughout this work (Fig. 1d,h). The low miscut sample ($\theta$ = 0.6°) exhibited mainly regular square-shaped hillocks (Fig. 1b,c) with smooth sidewalls and a slightly concave top surface (Fig. 1d). The average angle of the sidewall is ~23°, which is comparable to the 25.2° angle that a {113} face makes to the (001) plane. The junctions between the four top-faces of the hillock form a <100> cross where they meet. The inclination angle of the top surface is very small, in the range of 1–2° (see cross-sectional linecuts in SI Fig. S1.) While the CL intensity from the hillock-free background is dim, the entire hillock luminesces brighter with an additional band of stronger luminescence in the shape of a square loop (Fig. 1c). This inner square ring corresponds to the short nitrogen doping period as illustrated in the schematics in Fig. 1e,f and experimentally shown later in the paper in Fig. 3a,b. In contrast, the higher miscut sample ($\theta$=1.25°) exhibited much more irregularly shaped hillocks that were elongated along the direction perpendicular to step flow (see Fig. 1h,i). Additionally, the surface also exhibited step bunches throughout the sample such as the ones highlighted in Fig. 1h,i and later in Fig. 5. The hillocks themselves appear to have much rougher sidewalls (Fig. 1j) compared to the smoother sidewalls of the low miscut sample’s hillocks. These elongated hillocks also luminesce strongly, with bright inner traces of the hillock’s shape corresponding to the nitrogen doping period. We note that this square hillock morphology is not ubiquitous and appears sensitive to chamber history and/or surface preparation, suggesting it represents a narrow but revealing growth regime rather than a generic outcome of (001) diamond growth.

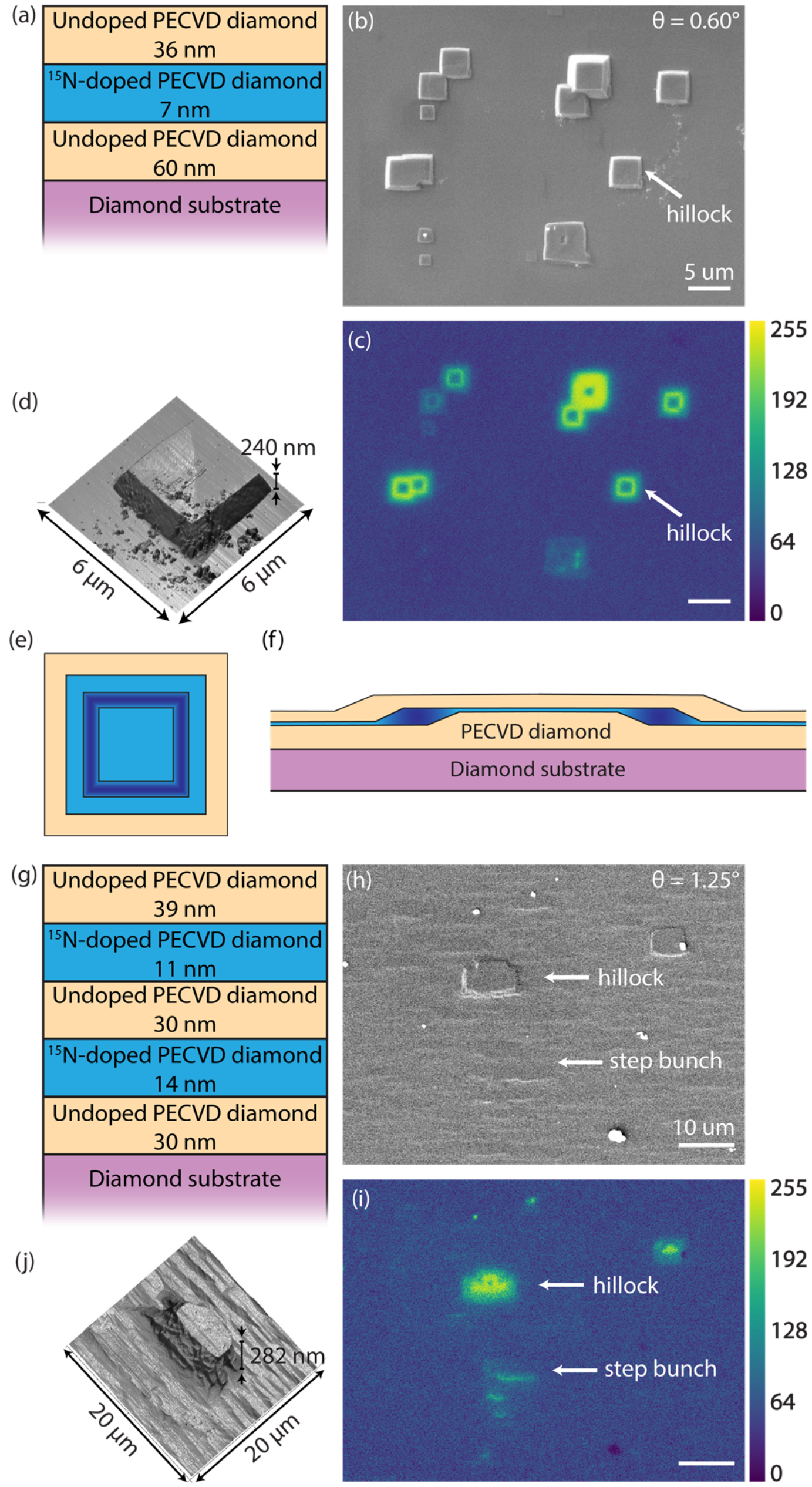

**Fig. 1:** (a) Growth schematic depicting layer thicknesses, (b) scanning electron microscope (SEM) secondary electron image and (c) panchromatic cathodoluminescence (CL) image of the same region showing multiple hillocks from the low miscut (θ = 0.6°) sample. The hillock pointed to in (b) and (c) is also represented via a (d) 3D atomic force microscopy (AFM) rendering. The color bar to the right of (c) is in grayscale value. (e) Top-down and (f) cross-sectional schematics of a classic hillock from the low miscut (θ = 0.6°) sample depicting why CL signal is observed in a ring around the center of the hillock, where blue represents nitrogen doping and dark blue represents increased nitrogen doping in hillock sidewalls. (g-j) Corresponding schematic, images, and hillock 3D rendering for the high miscut (θ = 1.25°) sample. Another surface feature, a step bunch, is also highlighted in (h) and (i).

Historically, pyramidal hillocks in diamond and other materials have been associated with growth spirals, nucleating from a screw dislocation at the center.[24–27] However, the anomalous inverted top of the hillocks in the low-miscut sample hints at a different origin. To investigate further, a TEM cross-section (Fig. 2) was prepared using a platinum (Pt) cross to mark the center of the hillock and ensure its preservation in the foil, see Fig. S2a in the Supplemental Information (SI). Fig. 2a shows a wide overview of the hillock in cross-section showing dark lines characteristic of dislocation and stacking fault strain contrast at the center. Dark trapezoidal regions due to strain contrast are also visible on either side of the hillock center, nominally at locations where we expect nitrogen doping based on the bands of enhanced CL—we substantiate this claim using spatially-resolved secondary ion mass spectrometry (nanoSIMS) subsequently. A [110] zone axis bright-field reconstruction of the center defective region was produced using 4D-STEM in Fig. 2b, demonstrating stacking faults nucleate from points slightly above the re-growth interface (seen as a dark horizontal line of contrast, labelled in Fig. 2a). Dashed lines extend to the regions from which local diffraction patterns were obtained and averaged along the stacking fault partials, as well as a defect-free region. The streakiness in the stacking fault's diffraction pattern is common for these extended defects[28] and confirms, along with the atomic resolution imaging shown in Fig. S2f in the SI, that these features are indeed stacking faults. Notably, we observe two threading dislocations arising from the peak of the stacking fault formation at the center of the hillock. The dislocations are visible at the g = 220 two-beam condition (two dark vertical lines in the center of the foil in Fig. 2d) and nearly disappear at the g = 004 two-beam condition (Fig. 2c), implying that these dislocations are predominantly edge-type rather than screw-type. The extended defects are likely roughening-induced from growth that is increasingly more disordered and rough at the center.[29,30] The thin dark line cutting horizontally across the foil is the thin film/substrate interface, which appears to bend towards the center on either side of the stacking fault-dense center. This implies there may have been some etch pit or divot that nucleated this disordered growth that led to stacking faults which convert eventually to threading edge dislocations.[31] The low angle inverted or dimpled surface of the hillock likely reflects growth kinetics, with a higher effective step density at the hillock periphery relative to the center. In contrast, the predominantly edge-type dislocations identified at the center of the hillock would not be expected to sustain spiral step generation in the absence of a significant screw component.

To our knowledge, direct visualization of nitrogen-dense regions in diamond via diffraction strain contrast in TEM has not been previously reported. The contrast band is observed reproducibly under both g = 004 and g = 220 two-beam conditions, consistent with diffraction contrast from a local lattice distortion associated with hydrostatic strain from nitrogen rather than g-vector-specific dislocation contrast.[32] While this contrast is not chemically informative, its depth and lateral extent coincide with the intentionally introduced nitrogen interval as established later. Where strain contrast was sufficiently sensitive, attempts to characterize these nitrogen-doped regions using chemically sensitive Energy Dispersive Spectroscopy (EDS) and Electron Energy Loss Spectroscopy (EELS) failed to pick up nitrogen signal above the carbon background. We later show the nitrogen concentration in the sidewall is below the practical detection limit of EDS/EELS for nitrogen in a carbon matrix under these conditions, ~0.1at% or $10^{20}$ $cm^{-3}$.[33] The nitrogen strain contrast acts as a marker that tracks subtle growth transitions from: 1) a higher hillock sidewall angle growth around 27° before the nitrogen doping occurred (the inner edge of the nitrogen-dense trapezoid) to 2) a lower sidewall angle at the outer edge of the nitrogen-doped region of 19° and back again to 3) a higher sidewall angle of 24° for the rest of the hillock growth. Observing the lower outer edge angle compared to the higher inner edge of the nitrogen trapezoid region demonstrates that the nitrogen doping itself increases the lateral to vertical growth ratio compared to the undoped hillock sidewall.

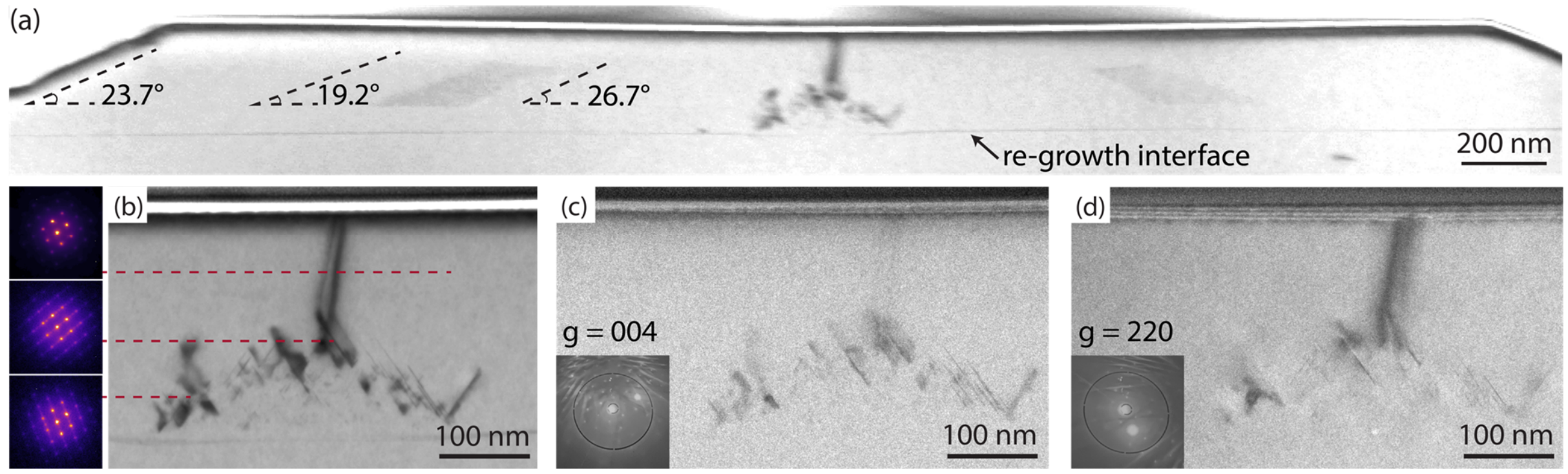

**Fig. 2:** (a) Overview cross-sectional bright-field (BF) 4D-STEM reconstruction of the hillock cross-section at the [110] zone axis (ZA). Stacking faults and dislocations are visible at the center of the hillock. The two trapezoidal areas of higher contrast left and right of center are the nitrogen-dense sidewall regions. The hillock sidewall angle, as well as the inner and outer sidewall angles of the nitrogen trapezoid are indicated via dashed black lines. The re-growth interface is pointed to with a black arrow. (b) BF reconstruction from 4D-STEM of the center defective region at the [110] ZA. Diffraction patterns from selected regions are indicated with burgundy dashed lines. (c,d) Higher magnification images of the extended defects at g = 004 and g = 220 two-beam conditions, respectively, with insets of the diffraction pattern at these imaging conditions.

We quantify the nitrogen incorporation at the defect-induced hillock sidewalls via site-coincident spatially resolved SIMS, CL, and AFM. Fig. 3a,b shows CL and corresponding depth-averaged $^{15}N$ SIMS isotope map from an exemplary hillock from the low-miscut sample. The region of higher CL is sequestered to an inner square loop in the hillock, which corresponds extremely well to the area of greater nitrogen density in the $^{15}N$ SIMS isotope map. NanoSIMS indicates a peak $^{15}N$ concentration of ~$2\times10^{19}$ $cm^{-3}$ within the square loop, 1000x higher than the film background and in line with the visibility of diffraction strain contrast in the 100-200 nm thick TEM foil. The 15N depth profiles show that the nitrogen-doped interval appears much broader in depth when probed through the inclined hillock sidewall compared to the planar background film, consistent with the greater path length through the angled sidewall geometry and TEM observations of a thicker nitrogen-doped layer in the sidewall. The hillock center itself also aggregates more nitrogen than the background film, albeit in 10× lower density than the sidewall. Greater incorporation of nitrogen as well as other dopants has been observed in (111) and (110)-oriented films compared to (001)-oriented films.[12,34–36] The higher incorporation of nitrogen in the hillock could be related to a similar mechanism, due to the sidewall being a distinct facet. The sidewall angles ranging from 19-25° are closest to the angle the {113} facet makes with the (001) surface, 25.2°.[37] Fig. 3d shows hyperspectral CL spectra from a typical hillock and the background film. The higher intensity and much more pronounced $NV^0$ peak in the hillock compared to the background film indicates that the greater luminescence in the hillock sidewalls is at least in part from nitrogen incorporating into NV-centers. As is common in CL experiments, we measure only $NV^0$ signal due to the electron beam converting $NV^-$ defects into $NV^0$, demonstrated via pump-probe spectroscopy by Solà-Garcia et al.[38] The presence of the H3 peak as well as the broad A-band peak indicate other point defects or complexes may also be present.[39–41]

It is widely accepted that nitrogen incorporates substitutionally in diamond and preferentially incorporates at step edges.[42] Naturally, we would expect a greater incorporation of nitrogen where the step edge density is greater, and we would also expect the hillock sidewall angle to correlate to step edge density. The process to obtain sidewall angle from AFM is detailed in the SI and Fig. S1. Correlated measurements (Fig. 3e) of average sidewall angle and $^{15}N$ density from nanoSIMS for three hillocks and thus twelve hillock sidewalls from the low-miscut sample demonstrate there is no observable dependence of nitrogen density on hillock sidewall angle. At the very least, the dependence is less than the maximum difference in nitrogen incorporation observed in the sidewalls divided by the maximum angle difference observed among sidewalls, $3.12 \times 10^{18}$ $cm^{-3}$ per degree miscut or 17.7 ppm per degree miscut. In (001)-oriented diamond films grown under Meynell et al.'s specific growth conditions, the dependence of nitrogen on miscut angle is ~10 ppm per degree miscut where nitrogen incorporation increases with increasing miscut.[42] While the impact of miscut on nitrogen incorporation in (113)-oriented films has not been studied, nitrogen incorporation in (111)-oriented films is independent of miscut.[12] Thus, the weak dependence of nitrogen incorporation on sidewall angle suggests that incorporation on the hillock sidewalls is not governed simply by (001)-type miscut and step-density scaling, and is more consistent with facet-specific incorporation kinetics. While we do not observe any impact of sidewall angle on nitrogen density, sidewall orientation appears to impact nitrogen aggregation. Nitrogen density in hillock sidewalls exhibits two-fold rather than the expected four-fold symmetry (Fig. 3f, S3d,h in the SI). We do not yet understand why this is and future systematic study of the impact of misorientation angle on nitrogen incorporation, as well as the impact of growth conditions and orientations on the hydrocarbon to nitrogen admolecule migration length (similar to Yamamoto et al.[43] for boron and phosphorus) is needed to reveal the mechanism of orientation-dependent incorporation.

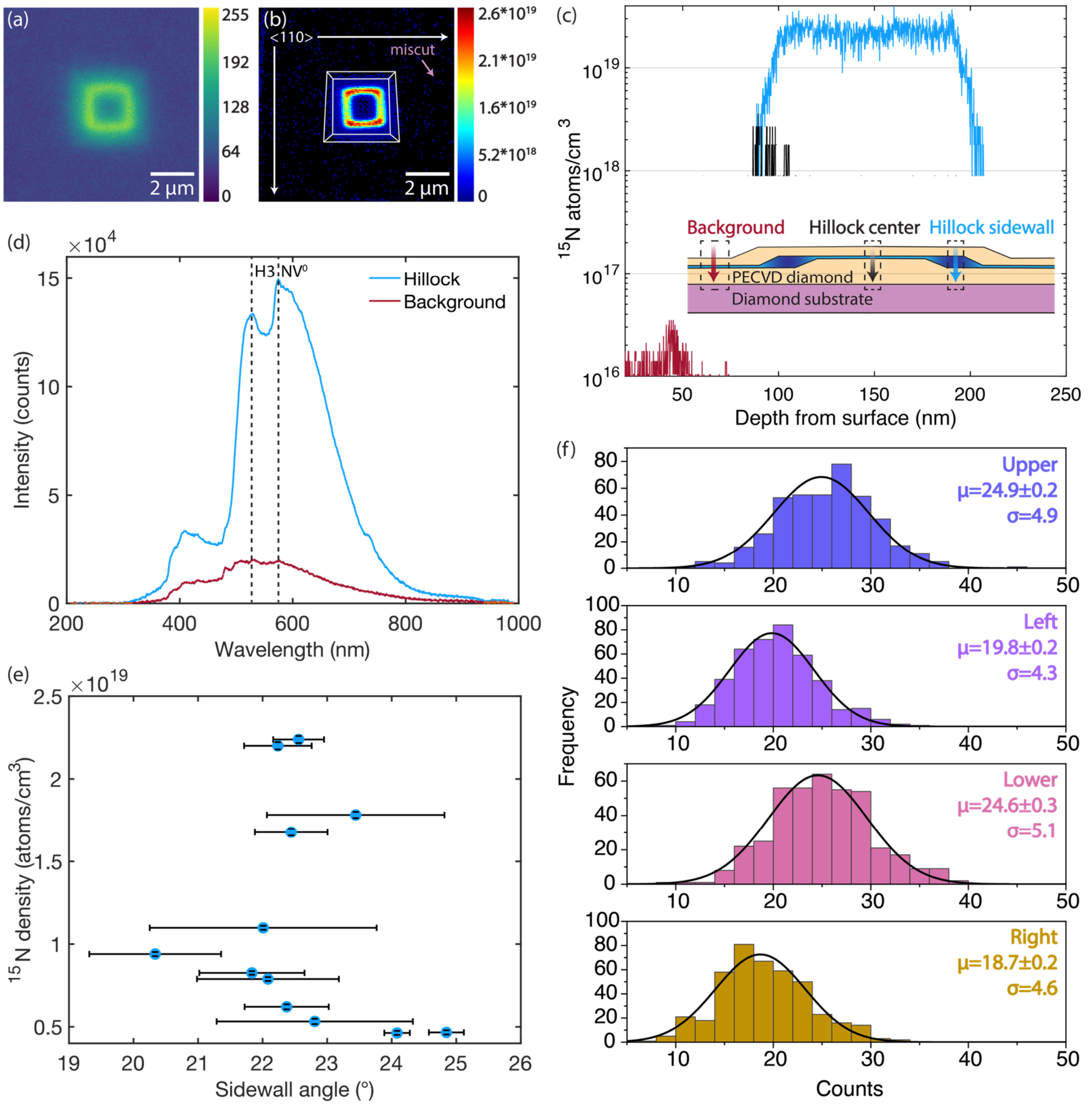

**Fig. 3:** (a) Panchromatic cathodoluminescence (CL) image of an exemplary hillock from the low (θ = 0.6°) miscut sample and (b) its corresponding $^{15}N$ nanoSIMS isotope map, demonstrating strong correlation between higher intensity luminescence and $^{15}N$ aggregation. Colorbars to the right of each image indicate grayscale value for the CL image (a) and $cm^{-3}$ for the isotope map (b). The white outline traces the overall shape of the hillock obtained from a secondary electron image. (c) $^{15}N$ depth profiles obtained from nanoSIMS in the background of the film (burgundy), hillock center (black), and hillock sidewall (blue) yield insight into the growth during doping, as depicted in the inset schematic. $^{15}N$ density is 1000× higher in the hillock sidewall compared to the background (planar) film. (d) Hyperspectral CL of a typical hillock compared to the background film shows that the greater sidewall luminescence originates at least in part from $NV^0$ centers. (e) $^{15}N$ density in hillock sidewalls from nanoSIMS is plotted vs hillock sidewall angle obtained from atomic force microscopy (AFM), demonstrating there is no observable dependence of $^{15}N$ density on sidewall angle. Vertical error bars represent standard error $^{15}N$ density and horizontal error bars represent standard deviation in angle measurement. (f) $^{15}N$ counts distributions among the hillock sidewalls from (b) show the lower and upper sidewalls contain higher 15N density than the left and right sidewalls, a common trend we observe among all hillocks surveyed in the low miscut sample.

Continuous-wave electron spin resonance (CWESR) was performed to interrogate the orientation of NV centers formed in the hillock sidewalls. Fig. 4a is a top-down CWESR map of a hillock, where the four different colors correspond to distinct NV orientations, as shown in Fig 4b. A different NV orientation is present at each sidewall, showing that despite the (001)-orientation of the film, the NVs are preferentially oriented. Linecuts taken along dashed white lines in Fig. 4a demonstrate that this preferential orientation is near-perfect. As the linecuts traverse vertically across the hillock passing through different sidewall domains, the CWESR signal remains tightly concentrated around one peak frequency for each sidewall domain. Aside from preferential orientation, other important metrics to qualify NV centers are the $T_2$ coherence time, the NV-center density, and the local substitutional nitrogen (P1) bath.[44] Fig. 4c shows two coherence decay curves of an exemplary hillock, one obtained from a Hahn echo sequence and one from a double electron-electron resonance (DEER) measurement targeting P1 centers. Three separate hillocks were characterized via these methods to obtain P1 density of the hillock sidewall following density extraction protocol from Hughes et al.[45] $T_2$ times ranged from 8.1–17.6 μs and P1 density ranged from 5.4–10.4 ppm, consistent with P1-limited decoherence.[46] The P1 density extracted from DEER is on the lower end of the measured nanoSIMS concentration of 10–100 ppm range in the hillock sidewalls, comparable to the difference in P1 and SIMS nitrogen density observed by Hughes et al. for delta-doped nitrogen layers.[45] We similarly attribute this difference to uncertainties and variability in the SIMS measurements and DEER. An NV decoherence-based method using XY-8 sequences[45] was used to obtain NV densities in the range of 0.11–0.19 ppm, demonstrating a 1.7–2% NV/P1 ratio. While (111)-oriented and (113)-oriented substrates have achieved even higher grown-in NV conversion of 4–11% and 25%, respectively, the 1.7–2% NV/P1 ratio in hillock sidewalls is still considerably higher than 0.5%, the grown-in P1-to-NV conversion for (100) and (110)-oriented thin films.[12,47,48] Though comparison of the hillock sidewalls and bulk (111) and (113)-oriented substrates is not direct, the lower NV-conversion efficiency of the hillock sidewalls may point to a mixed growth geometry.

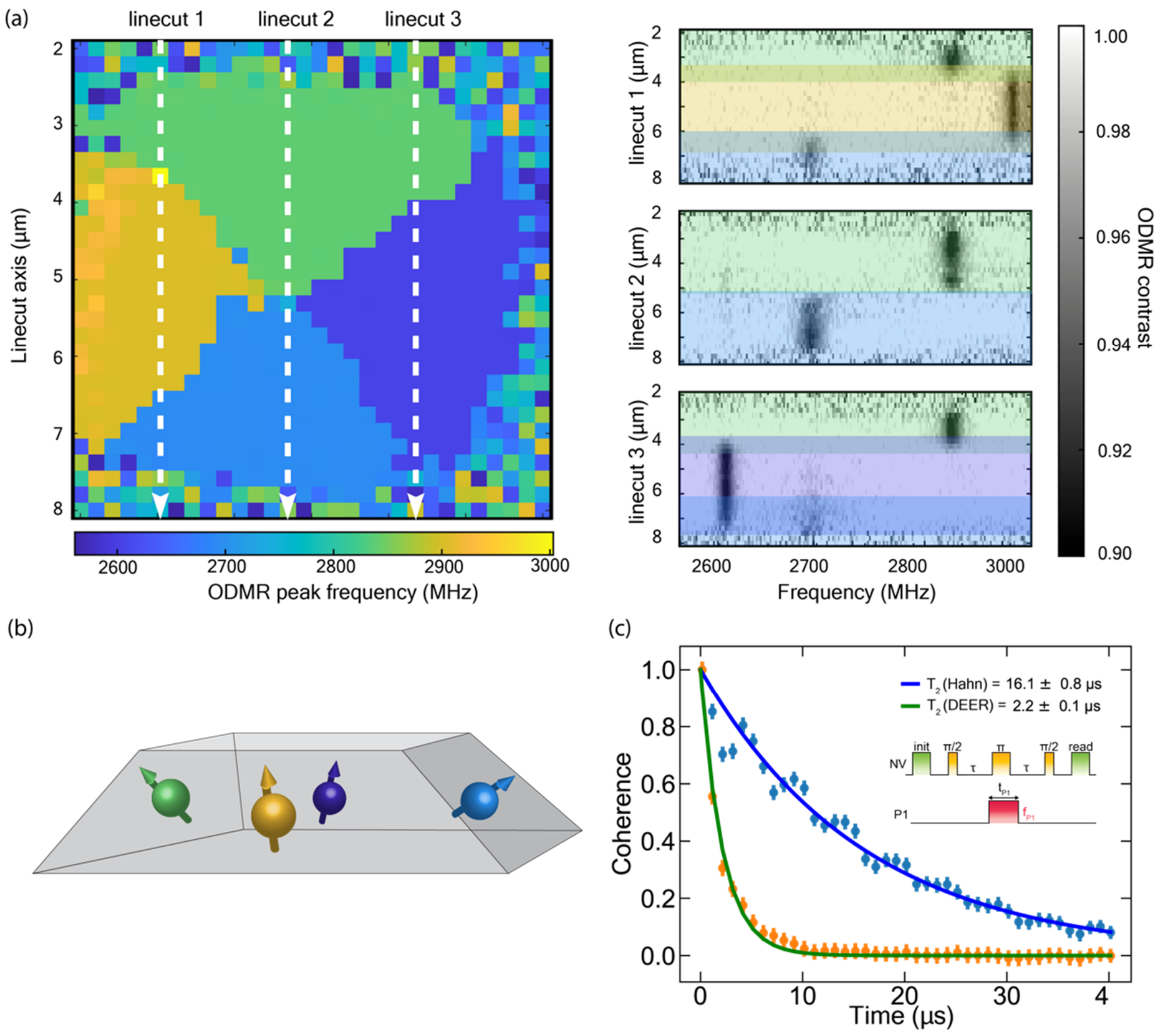

**Fig. 4**: (a) CWESR frequency map detailing a different preferential orientation of NVs on each separate hillock sidewall as shown in the (b) schematic below. Linecuts demonstrate preferential orientation on the sidewalls is near-perfect, with ODMR contrast isolated to the individual peak frequency for each independent sidewall. (c) Hahn echo sequence and DEER on one exemplary hillock from which $T_2$ times are obtained. A diagram of the pulse sequence used for the measurement is included as an inset.

The preferential orientation we observe in the hillock sidewalls supports our hypothesis that the sidewalls are forming either a distinct {113} facet or contain a combination of {001} and {111} facets, as NV preferential orientation tends to align with the crystal growth direction and a different growth direction exists for each of the four sidewalls.[49] Prior work on grown-in NVs on true (113)-oriented substrates shows the predominant orientation is angled closest to the growth direction.[48,49] Yet, there are still two other minority NV orientations detected in those reports that we do not see. Near-perfect single preferential orientation of grown-in NVs has most commonly been achieved on (111)-oriented surfaces.[21] As shown in Fig. 4a,b, we observe a single orientation throughout each hillock quadrant. Note that we define quadrant as the entirety of the hillock segment that originates from the center and grows out in the {113} facet direction. Therefore, this observation introduces two important discussion points: 1) that NVs are preferentially-oriented throughout the entirety of the hillock quadrant, not just within the intentional nitrogen-doped growth period (inner square ring in Fig. 1c and Fig. 3a,b) and 2) that the NVs exhibit a single preferential orientation within each hillock quadrant, which is expected of a {111} facet but not of the {113} facet. To address the first point, we rationalize that as the hillock grows outwards, it continuously incorporates preferentially oriented NVs at its sidewalls during the nitrogen-doping growth period as well as before and after from residual nitrogen in the background of the chamber. Yet, as the hillock grows and during the nitrogen-doping period, its top surface which is inclined much closer to the (001)-surface must be incorporating NVs in all four possible orientations. Here, we invoke the much higher nitrogen incorporation and NV-conversion efficiency of the sidewalls ($2\times10^{19}$ $cm^{-3}$ of $^{15}N$ and 1.7-2% NV-conversion efficiency) in our hillocks compared to {001} facets ($2\times10^{18}$ $cm^{-3}$ of $^{15}N$ and 0.5% NV-conversion efficiency).[47] Applying the NV incorporation efficiencies to the nitrogen densities observed from nanoSIMS in the hillock center and sidewalls (Fig. 3c), we find that the NV incorporation on the {001} plane would account for only ~2-3% of the total NVs formed within the hillock quadrants. Thus, CWESR measures near-perfect preferential orientation throughout the four distinct hillock quadrants.

To address the second point, the single orientation despite the {113} facet angle can be attributed to the particular growth conditions in this sample or the sidewalls containing a combination of {001} and {111} facets. Chouaieb et al. have shown that decreasing the growth temperature from 1000°C to 800°C results in increasing the amount of NVs oriented out-of-plane from 47.5% to 79% on [113]-oriented substrates.[50] These growth conditions may favor rapid lateral step nucleation together with limited carbon adatom mobility, increasing the likelihood that a vacancy forms directly above nitrogen.[48] While our sample was also grown at 800°C, its near-100% preferential orientation could similarly be explained by kinetic growth parameters such as gas flow and substrate miscut. Atomistic modeling of the {113} preferential orientation mechanism as well as further investigation into the impact of growth conditions on preferential orientation in hillocks would clarify this mechanism.[51] The predominantly single-oriented NVs could also be a result of the sidewalls containing a combination of {111} and {001} facets, where again the {111} facets exhibit a much higher NV-conversion efficiency compared to {001}. Neither AFM nor XS-STEM showed the distinct 54.7° and 0°-angled sidewall steps we would expect of a staircase-like structure derived from multiple facets. AFM was performed at ambient pressure and thus, the resolution may not have been sufficient to observe these facets. Projection from the thickness of the XS-STEM foil may have also limited the resolution able to observe this structure. Other examples in the literature exist of facets that are far from {111} exhibiting a dominant NV-orientation,[52,53] though until now, only facets close to {111} exhibited predominantly a single NV-orientation, and these are all engineered via overgrowth of etched structures.[54–57]

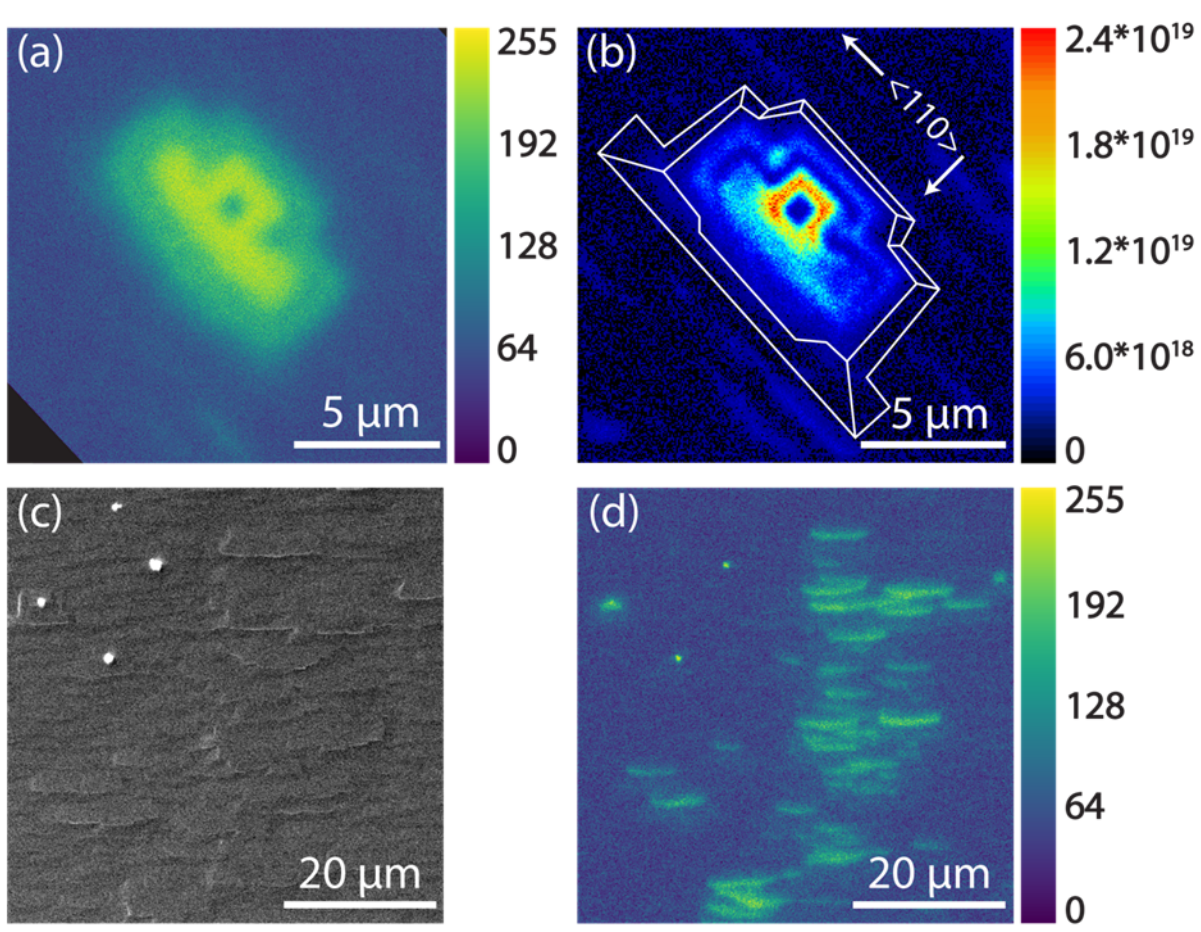

**Fig. 5**: High miscut (θ = 1.25°) sample hillock's (a) panchromatic cathodoluminescence image with colorbar to the right representing grayscale value and (b) corresponding nanoSIMS $^{15}N$ isotope map, with white outlines drawn in from overall hillock shaped observed in the secondary electron image. The colorbar to the right is in units of $^{15}N$ $cm^{-3}$. (c) Secondary electron image of a step-bunch region in the high miscut (θ=1.25°) sample and (d) panchromatic CL image of the same region with colorbar to the right in grayscale value.

The near-perfect preferential orientation and up to 1000× higher nitrogen incorporation in the hillock sidewalls from the low miscut sample (θ = 0.60°) naturally points to investigating surface features present at higher miscuts. As seen earlier in Fig. 2, the high miscut (θ=1.25°) sample's surface features, both hillocks and step bunches, also experience heightened CL signal compared to the background of the film. Fig. 5a,b focus on a hillock from the high miscut sample and its corresponding nanoSIMS $^{15}N$ isotope map. The two delta-doped $^{15}N$ layers in the higher miscut sample result in two dense nitrogen-containing rings, higher in nitrogen content than the background of the film. The first delta-doped layer resulted in a higher density nitrogen region compared to the surrounding ring resulting from the second delta-doped layer. The two nitrogen-dense regions allow us to observe how the hillock's shape evolved between the nitrogen-doped layers, growing wider at a slower rate compared to the nitrogen-doping period. Step bunches that have aggregated 10× more nitrogen compared to the background are also visible as faint striations surrounding the hillock in Fig. 5b. Fig. 5c,d demonstrate how a larger step-bunch feature experiences CL signal as strong as the hillock center, suggesting that step bunches can also strongly aggregate nitrogen. This indicates that step bunches are relevant surface features for understanding how lower-dimensional surface morphology influences nitrogen aggregation and potentially NV-center formation. Interestingly, the nitrogen aggregation is capped around $2\times10^{19}$ $cm^{-3}$ in the surface features both in the high and low miscut sample, regardless of background doping, suggesting perhaps a maximum incorporation efficiency for {113}-oriented facets.

In this work, we employ correlative microscopies and spatially resolved spectroscopy to examine nitrogen incorporation and NV-center formation at surface nonuniformities in (001)-oriented diamond. We demonstrate that hillocks, surface features typically considered undesirable for diamond quantum devices, locally aggregate nitrogen at concentrations up to three orders of magnitude above the surrounding film and give rise to enhanced NV-center density. Despite edge dislocations and stacking faults present at the center of the hillock associated with their growth, NV-centers in the hillock sidewall regions still exhibit P1-limited coherence. Thus, spontaneously occurring crystal facets oriented at 19–25° to the (001) surface are associated with enhanced nitrogen incorporation and elevated grown-in NV/P1 ratios locally in these structures. Importantly, we observe that in these hillocks, neither pyramidal overgrowth nor perfectly {111}-angled sidewalls are necessary to obtain near-single-orientation NV-centers at each distinct hillock sidewall. While the mechanism underlying the single-orientation NV populations observed at each hillock sidewall is not yet well understood, the facet-resolved growth appears highly ordered.

Our results motivate atomistic simulations and controlled overgrowth studies spanning facet inclinations from near-(001) toward the {113} and {111} orientations to further clarify the mechanism underlying preferential alignment. We also highlight step bunches as additional surface features of interest for understanding NV-center ensemble dimensionality and preferential orientation. Overall, these findings reveal spontaneous hillocks and related nonuniformities as valuable local probes and natural laboratories for understanding nitrogen aggregation and facet-dependent NV orientation in (001) diamond.

**Data Availability Statement:** The data that support the findings of this study are available from the corresponding authors upon reasonable request.

**Supporting Information**: Additional experimental details, including growth and characterization methodology; details of TEM sample preparation and additional imaging conditions for the nitrogen contrast and stacking faults within the hillock; additional evidence of the impact of sidewall orientation on nitrogen density in hillocks of varying dimensions.

**Acknowledgments:**

Diamond materials characterization (TEM, SEM, CL, AFM, nanoSIMS) was funded by DOE Award No. DE-SC0019241. Diamond materials synthesis was supported by NSF Award No. DMR-1906325. Part of this work was performed at nano@stanford RRID:SCR_026695. We gratefully acknowledge Dr. Dale Burns and the Stanford Doerr School of Sustainability's Microchemical Analysis Facility for help with the hyperspectral and panchromatic CL. L. B. H. W. acknowledges support from the NSF Graduate Research Fellowship Program (Grant No. DGE 2139319) and the UCSB Quantum Foundry. S. A. M. acknowledges support from Canada NSERC (Grant No. AID 516704- 2018) and the UCSB Quantum Foundry.

**References**

1. Aslam, N. *et al.* Nanoscale nuclear magnetic resonance with chemical resolution. *Science (80-. ).* **357**, 67–71 (2017).
2. Rondin, L. *et al.* Magnetometry with nitrogen-vacancy defects in diamond. *Reports Prog. Phys.* **77**, (2014).
3. Cai, J., Retzker, A., Jelezko, F. & Plenio, M. B. A large-scale quantum simulator on a diamond surface at room temperature. *Nat. Phys.* **9**, 168–173 (2013).
4. Rodgers, L. V. H. *et al.* Materials challenges for quantum technologies based on color centers in diamond. *MRS Bull.* **46**, 623–633 (2021).
5. Pfaff, W. *et al.* Unconditional quantum teleportation between distant solid-state quantum

bits. *Science (80-. ).* **345**, 532–535 (2014).

6. Chatterjee, A. *et al.* Semiconductor qubits in practice. *Nat. Rev. Phys.* **3**, 157–177 (2021).
7. Balasubramanian, G. *et al.* Ultralong spin coherence time in isotopically engineered diamond. *Nat. Mater.* **8**, 383–387 (2009).
8. Degen, C. L., Reinhard, F. & Cappellaro, P. Quantum sensing. *Rev. Mod. Phys.* **89**, (2017).
9. Taylor, J. M. *et al.* High-sensitivity diamond magnetometer with nanoscale resolution. *Nat. Phys.* **4**, 810–816 (2008).
10. Acosta, V. M. *et al.* Diamonds with a high density of nitrogen-vacancy centers for magnetometry applications. *Phys. Rev. B - Condens. Matter Mater. Phys.* **80**, (2009).
11. Pham, L. M. *et al.* Enhanced metrology using preferential orientation of nitrogen-vacancy centers in diamond. *Phys. Rev. B - Condens. Matter Mater. Phys.* **86**, (2012).
12. Hughes, L. B. *et al.* Strongly Interacting, Two-Dimensional, Dipolar Spin Ensembles in (111)-Oriented Diamond. *Phys. Rev. X* **15**, (2025).
13. Choi, S., Yao, N. Y. & Lukin, M. D. Quantum metrology based on strongly correlated matter. (2017).
14. Zhou, H. *et al.* Quantum Metrology with Strongly Interacting Spin Systems. *Phys. Rev. X* **10**, (2020).
15. Chakraborty, T. *et al.* CVD growth of ultrapure diamond, generation of NV centers by ion implantation, and their spectroscopic characterization for quantum technological applications. *Phys. Rev. Mater.* **3**, (2019).
16. Eichhorn, T. R., Mclellan, C. A. & Bleszynski Jayich, A. C. Optimizing the formation of depth-confined nitrogen vacancy center spin ensembles in diamond for quantum sensing. *Phys. Rev. Mater.* **3**, (2019).
17. Tallaire, A. *et al.* High NV density in a pink CVD diamond grown with N2O addition. *Carbon N. Y.* **170**, 421–429 (2020).
18. Kollarics, S. *et al.* Ultrahigh nitrogen-vacancy center concentration in diamond. *Carbon N. Y.* **188**, 393–400 (2022).
19. Ozawa, H., Ishiwata, H., Hatano, M. & Iwasaki, T. Thermal Stability of Perfectly Aligned Nitrogen-Vacancy Centers for High Sensitive Magnetometers. *Phys. Status Solidi Appl. Mater. Sci.* **215**, (2018).

20. Teraji, T. *et al.* Homoepitaxial diamond film growth: High purity, high crystalline quality, isotopic enrichment, and single color center formation (Phys. Status Solidi A 11⁄2015). *Phys. Status Solidi Appl. Mater. Sci.* **212**, (2015).
21. Michl, J. *et al.* Perfect alignment and preferential orientation of nitrogen-vacancy centers during chemical vapor deposition diamond growth on (111) surfaces. *Appl. Phys. Lett.* **104**, (2014).
22. Tallaire, A., Kasu, M., Ueda, K. & Makimoto, T. Origin of growth defects in CVD diamond epitaxial films. *Diam. Relat. Mater.* **17**, 60–65 (2008).
23. Blinder, R. *et al.* Reducing inhomogeneous broadening of spin and optical transitions of nitrogen-vacancy centers in high-pressure, high-temperature diamond. *Commun. Mater.* **5**, (2024).
24. Tokuda, N. Homoepitaxial diamond growth by plasma-enhanced chemical vapor deposition. in *Novel aspects of diamond: from growth to applications* 1–29 (2014).
25. Kawanishi, S., Kamiko, M., Yoshikawa, T., Mitsuda, Y. & Morita, K. Analysis of the Spiral Step Structure and the Initial Solution Growth Behavior of SiC by Real-Time Observation of the Growth Interface. *Cryst. Growth Des.* **16**, 4822–4830 (2016).
26. Zhou, K. *et al.* Hillock formation and suppression on c-plane homoepitaxial GaN Layers grown by metalorganic vapor phase epitaxy. *J. Cryst. Growth* **371**, 7–10 (2013).
27. Lin, C. T. Study of growth spirals and screw dislocations on YBa2Cu3O7-δ single crystals. *Phys. C Supercond. its Appl.* **337**, 312–316 (2000).
28. Steeds, J. W., Gilmore, A., Wilson, J. A. & Butler, J. E. On the nature of extended defects in CVD diamond and the origin of compressive stresses. *Diam. Relat. Mater.* **7**, 1437–1450 (1998).
29. Dong, L., Schnitker, J., Smith, R. W. & Srolovitz, D. J. Stress relaxation and misfit dislocation nucleation in the growth of misfitting films: A molecular dynamics simulation study. *J. Appl. Phys.* **83**, 217–227 (1998).
30. Yan, H. *et al.* Multi-microscopy characterization of threading dislocations in CVD-grown diamond films. *Appl. Phys. Lett.* **124**, (2024).
31. Tsubouchi, N. Conversions of a stacking fault to threading dislocations in homoepitaxial diamond growth studied by transmission electron microscopy. *Appl. Phys. Lett.* **117**, (2020).

32. Stoupin, S. & Shvyd'Ko, Y. V. Ultraprecise studies of the thermal expansion coefficient of diamond using backscattering x-ray diffraction. *Phys. Rev. B - Condens. Matter Mater. Phys.* **83**, (2011).
33. Hudak, B. M. & Stroud, R. M. Atomically Precise Detection and Manipulation of Nitrogen-Vacancy Centers in Nanodiamonds. *ACS Nano* **17**, 7241–7249 (2023).
34. Samlenski, R. *et al.* Incorporation of nitrogen in chemical vapor deposition diamond. *Appl. Phys. Lett.* **67**, 2798 (1995).
35. Kato, H., Makino, T., Yamasaki, S. & Okushi, H. N-type diamond growth by phosphorus doping on (0 0 1)-oriented surface. *J. Phys. D. Appl. Phys.* **40**, 6189–6200 (2007).
36. Mortet, V. *et al.* Effect of substrate crystalline orientation on boron-doped homoepitaxial diamond growth. *Diam. Relat. Mater.* **122**, (2022).
37. Janssen, G., Schermer, J. J., van Enckevort, W. J. P. & Giling, L. J. On the occurrence of {113}; facets on CVD-grown diamond. *J. Cryst. Growth* **125**, 42–50 (1992).
38. Solà-Garcia, M., Meuret, S., Coenen, T. & Polman, A. Electron-Induced State Conversion in Diamond NV Centers Measured with Pump-Probe Cathodoluminescence Spectroscopy. *ACS Photonics* **7**, 232–240 (2020).
39. Zaitsev, A. *Optical Properties of Diamond*. (Springer, 2001).
40. Collins, A. T., Kamo, M. & Sato, Y. A spectroscopic study of optical centers in diamond grown by microwave-assisted chemical vapor deposition. *J. Mater. Res.* **5**, 2507–2514 (1990).
41. Collins, A. T. The characterisation of point defects in diamond by luminescence spectroscopy. *Diam. Relat. Mater.* **1**, 457–469 (1992).
42. Meynell, S. A. *et al.* Engineering quantum-coherent defects: The role of substrate miscut in chemical vapor deposition diamond growth. *Appl. Phys. Lett.* **117**, (2020).
43. Yamamoto, T., Janssens, S. D., Ohtani, R., Takeuchi, D. & Koizumi, S. Toward highly conductive n -type diamond: Incremental phosphorus-donor concentrations assisted by surface migration of admolecules. *Appl. Phys. Lett.* **109**, (2016).
44. Trusheim, M. E. *et al.* Scalable fabrication of high purity diamond nanocrystals with long-spin-coherence nitrogen vacancy centers. *Nano Lett.* **14**, 32–36 (2014).
45. Hughes, L. B. *et al.* Two-dimensional spin systems in PECVD-grown diamond with tunable density and long coherence for enhanced quantum sensing and simulation. *APL*

*Mater.* **11**, (2023).
46. Bauch, E. *et al.* Decoherence of ensembles of nitrogen-vacancy centers in diamond. *Phys. Rev. B* **102**, (2020).
47. Edmonds, A. M. *et al.* Production of oriented nitrogen-vacancy color centers in synthetic diamond. *Phys. Rev. B - Condens. Matter Mater. Phys.* **86**, (2012).
48. Balasubramanian, P. *et al.* Enhancement of the creation yield of NV ensembles in a chemically vapour deposited diamond. *Carbon N. Y.* **194**, 282–289 (2022).
49. Lesik, M. *et al.* Preferential orientation of NV defects in CVD diamond films grown on (113)-oriented substrates. *Diam. Relat. Mater.* **56**, 47–53 (2015).
50. Chouaieb, S. *et al.* Optimizing synthetic diamond samples for quantum sensing technologies by tuning the growth temperature. *Diam. Relat. Mater.* **96**, 85–89 (2019).
51. Miyazaki, T. *et al.* Atomistic mechanism of perfect alignment of nitrogen-vacancy centers in diamond. *Appl. Phys. Lett.* **105**, (2014).
52. Weippert, J. *et al.* NV-doped microstructures with preferential orientation by growth on heteroepitaxial diamond. *J. Appl. Phys.* **133**, (2023).
53. Lang, N. *et al.* Controlled lateral positioning of NV centres in diamond by CVD overgrowth. *Phys. Scr.* **99**, (2024).
54. Götze, A. *et al.* Preferential Placement of Aligned Nitrogen Vacancy Centers in Chemical Vapor Deposition Overgrown Diamond Microstructures. *Phys. Status Solidi - Rapid Res. Lett.* **16**, (2022).
55. Lebedev, V. *et al.* Epitaxial Lateral Overgrowth of Wafer-Scale Heteroepitaxial Diamond for Quantum Applications. *Phys. Status Solidi Appl. Mater. Sci.* **221**, (2024).
56. Batzer, M. *et al.* Single crystal diamond pyramids for applications in nanoscale quantum sensing. *Opt. Mater. Express* **10**, 492 (2020).
57. Engels, J. *et al.* High ODMR contrast and alignment of NV centers in microstructures grown on heteroepitaxial diamonds. *Appl. Phys. Lett.* **124**, (2024).

**Supporting Information**

# I. Methodology

*a. Thin film growth and post-processing*

Diamond homoepitaxial growth and nitrogen doping were performed via plasma-enhanced chemical vapor deposition (PECVD) using a SEKI SDS6300 reactor on (001)-oriented electronic grade diamond substrates (Element Six Ltd.). Prior to growth, the substrates were fine-polished by Syntek Ltd. to a surface roughness of ~200-300pm, followed by a ~500nm etch to relieve polishing-induced strain. The miscut was characterized using x-ray diffractometry (XRD) rocking curve measurements about the (004) omega peak. The growth conditions for the low-miscut sample consisted of a 750 W plasma containing 0.025% $^{12}CH_4$ in 400 sccm $H_2$ flow held at 25 torr and ~800°C according to a graphite heating stage setting. A 3 hour isotopically purified (99.98% $^{12}C$) buffer layer was grown, followed by a $^{15}N$-doped layer (5 sccm $^{15}N_2$ gas, doped for 3 hours), and finished with a 3 hour $^{12}C$ capping layer. The high-miscut sample was grown with similar conditions except that the buffer/capping layers were grown for 2 hours, and the growth structure contained two $^{15}N$-doped layers (5 sccm $^{15}N_2$ gas, 2 hours) where the first layer was doped at 800°C and the second at 900°C. After growth, the samples were characterized with secondary ion mass spectrometry (SIMS) depth profiling to estimate the isotopic purity, epilayer thickness, and properties of the nitrogen-doped layers. The low-miscut sample went through further electron irradiation and annealing treatments to generate enhanced NV center concentrations in localized areas, though these irradiated regions were not further studied in this work. All NV measurements reported in this work were performed on regions that were not intentionally electron-irradiated for vacancy creation. Irradiation was performed with the 200 keV electrons of TEM both at room temperature and while annealing at ~750°C in-situ using a specialized holder while maintaining the TEM vacuum at ~3.5e-5 Pa. This process was repeated a second time, subjecting the sample to two brief annealing periods. After irradiation and annealing, the sample was cleaned in a boiling triacid solution (1:1:1 $H_2SO_4$:$HNO_3$:$HClO_4$) and annealed in air at 450°C to oxygen terminate the surface and help stabilize the negative $NV^-$ charge state for further measurements.

*b. Cathodoluminescence*

Panchromatic cathodoluminescence (CL) was conducted in a JEOL JSM-IT500HR SEM at 5kV accelerating voltage and contrast 10.3. The CL background was captured by blanking the beam and capturing an image using the same imaging conditions. This background was subtracted in ImageJ from every CL image reported throughout this work. Hyperspectral CL was conducted in a JEOL JXA-8230 "SuperProbe" electron microprobe at 5kV accelerating voltage and 40nA beam current.

*c. Spatially resolved secondary ion mass spectrometry (nanoSIMS)*

A Cameca NanoSIMS 50L was used to spatially resolve $^{15}N$ concentration (measured as the $^{12}C^{15}N^{-}$ ion). The spatial resolution was 50-70nm using a 10 pA $Cs^{+}$ ion beam and a mass resolution greater than 9000 to resolve nearby isobaric interferences. The nanoSIMS $^{15}N$ signal was converted to counts/cm$^3$ by dividing the $^{15}N$ counts summed in a background region outside of the hillock, by the product of the depth and area from which the background was summed. A counts to atoms normalization factor was obtained by comparing the counts/cm$^3$ value to the atoms/cm$^3$ $^{15}N$ peak obtained from prior SIMS. We performed four separate nanoSIMS scans encompassing six different hillocks – in three different hillock regions in the low miscut sample and one hillock region for the higher miscut sample. Each scan was performed with the same beam current, though the pixel size differed between the low and high miscut sample's scans – the low miscut sample's scans (all square-shaped hillocks in Fig. 3 and S3) had pixel size of 78x78nm, while the high miscut sample's scan (Fig. 5) had pixel size of 56x56nm. The number of pixels (and thus, pixel size) was chosen to optimize scan time and resolution given the necessary scan size to encompass each hillock region. Each sample was tuned separately and re-tuning was performed prior to each scan to account for differences in sample tilt in different scan regions. Thus, background normalization was conducted separately within each scan, taking a background region from each scan region to convert from counts/cm$^3$ to atoms/cm$^3$.

The nitrogen density in specific hillock sidewalls as plotted in Fig. 3e was calculated for each four sidewalls within the hillock depicted in Fig. 3a,b as well as the top hillocks in the regions depicted in Fig. S3. The lower hillocks in these regions were not included in the plot as their sidewalls' nitrogen accumulation was impacted either by proximity to another hillock or impingement on another hillock. The sidewall density was calculated from a small region-of-interest (ROI), 20x1 pixels, in the most nitrogen-dense part of each hillock sidewall as observed in the $^{15}N$ isotope summation maps (Fig. 3b, Fig. 5b, S3c,g). The total nitrogen counts were summed over a representative section of the nitrogen depth profile (where the nitrogen counts were nominally constant), and then divided by the volume of the analyzed ROI to obtain average nitrogen counts/cm$^3$ within each sidewall. This value was then converted to atoms/cm$^3$ as described earlier and plotted in Fig. 3e with vertical error bars representing plus or minus one standard error of the mean. Standard error was chosen rather than standard deviation as each milled frame in nanoSIMS represents one average density measurement among the pixels in each frame. Consequently, the nitrogen density we report in Fig. 3e is an average of averages, so the standard error of the mean is an appropriate measure of measurement error.[1]

*d. Atomic force microscopy (AFM) for sidewall angle determination*

AFM was performed on the Park NX-10 with tapping mode using NSC15 probes. We analyzed sidewall angles only from the lower miscut sample that exhibited classic square-shaped hillocks (Fig. 3, S1). The hillock sidewall angle was calculated by averaging the angle from three AFM line profiles from within the nitrogen-dense sidewall region (note how in Fig. 3b the nitrogen-dense region does not extend to the corners of the full hillock as outlined in white). The horizontal error bars in Fig. 3e represent plus or minus one standard deviation from the average, calculated from these three line profiles for each hillock sidewall. The sidewall angles were obtained from each line profile by first thresholding the data to remove high and low points that could interfere with the linear regression performed to segment the profile into three pieces, separated by two breakpoints (Fig. S1b) using Okazaki's function for a piecewise linear model.[2] The sidewall angles were then calculated from the first and last segment of each vertical and horizontal profile. The matlab code used to calculate the sidewall angles is included here, where leading edge (LE) refers to the sidewall where the slope is increasing in the profile and the trailing edge (TE) refers to the sidewall where the slope is decreasing:

```
filename = '062425-R1H7vprofiles.txt';
delimiterIn = ' ';
headerlinesIn = 3;
A = importdata(filename,delimiterIn,headerlinesIn);
%for 3 profiles
h1=[A.data(:,1),A.data(:,2)];
h2=[A.data(:,3),A.data(:,4)];
h3=[A.data(:,5),A.data(:,6)];
%thresholding
h1thresh = h1(h1(:,2) >= 3*(10^-8)&h1(:,2) <= 5*(10^-7), :);
```

```
h2thresh = h2(h2(:,2) >= 3*(10^-8)&h2(:,2) <= 5*(10^-7), :);
h3thresh = h3(h3(:,2) >= 3*(10^-8)&h3(:,2) <= 5*(10^-7), :);
%plot
hold on;
plot(h1(:,1),h1(:,2),'--','DisplayName','H1');
plot(h2(:,1),h2(:,2),'--','DisplayName','H2');
plot(h3(:,1),h3(:,2),'--','DisplayName','H3');
plot(h1thresh(:,1),h1thresh(:,2),'-','DisplayName','H1 Thresh');
plot(h2thresh(:,1),h2thresh(:,2),'-','DisplayName','H2 Thresh');
plot(h3thresh(:,1),h3thresh(:,2),'-','DisplayName','H3 Thresh');
%segmented linear regression
[coef1,breakPt1,R21] = piecewiselm(h1thresh(:,1),h1thresh(:,2),3)
[coef2,breakPt2,R22] = piecewiselm(h2thresh(:,1),h2thresh(:,2),3)
[coef3,breakPt3,R23] = piecewiselm(h3thresh(:,1),h3thresh(:,2),3)
%plot linear regression breakpoints
scatter([breakPt1(1,1);breakPt1(1,3)],[breakPt1(1,2);breakPt1(1,4)],'o','DisplayName','BreakPt 1');
scatter([breakPt2(1,1);breakPt2(1,3)],[breakPt2(1,2);breakPt2(1,4)],'o','DisplayName','BreakPt 2');
scatter([breakPt3(1,1);breakPt3(1,3)],[breakPt3(1,2);breakPt3(1,4)],'o','DisplayName','BreakPt 3');
%average leading edge (LE) and trailing edge (TE) angle and stdev
TE1 = atan(coef1(1,5))*180/pi
TE2 = atan(coef2(1,5))*180/pi
TE3 = atan(coef3(1,5))*180/pi
LE1 = atan(coef1(1,1))*180/pi
LE2 = atan(coef2(1,1))*180/pi
LE3 = atan(coef3(1,1))*180/pi
averageLE3p1 = atan(mean([coef1(1,1);coef2(1,1);coef3(1,1)]))*180/pi
stdLE3p1 = std([atan(coef1(1,1))*180/pi;atan(coef2(1,1))*180/pi;atan(coef3(1,1))*180/pi])
averageTE3p1 = atan(mean([coef1(1,5);coef2(1,5);coef3(1,5)]))*180/pi
stdTE3p1 = std([atan(coef1(1,5))*180/pi;atan(coef2(1,5))*180/pi;atan(coef3(1,5))*180/pi])
```

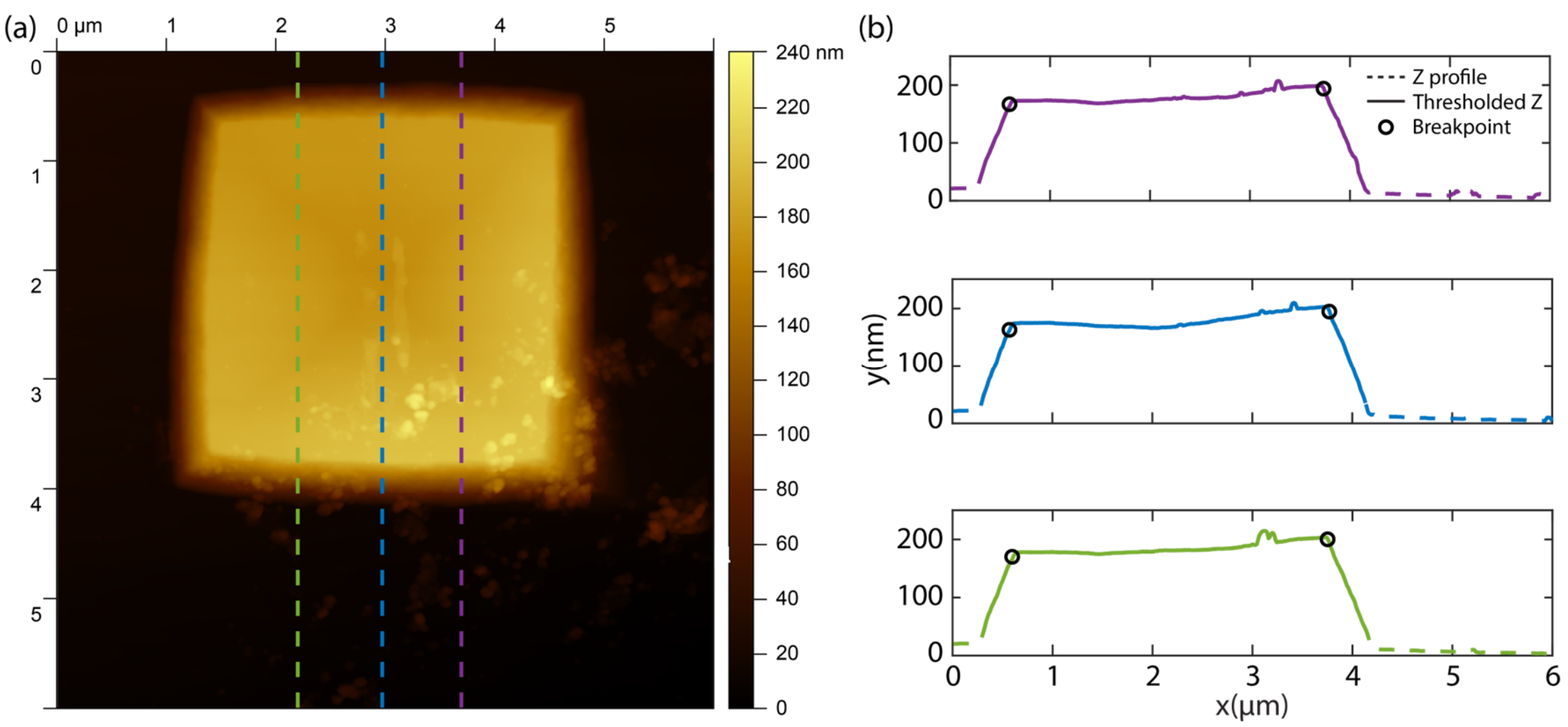

**Fig. S1:** (a) AFM image of the hillock analyzed in Fig. 3(a-c,f). Dashed green, blue, and purple lines correspond to the vertical Z profiles plotted in (b). Solid lines are thresholded Z profiles used to calculate sidewall angles from line segments split by breakpoints (solid circles).

### e. *Scanning and transmission electron microscopy (S-TEM) sample preparation and imaging*

An FEI Helios NanoLab 600i DualBeam focused ion beam (FIB) was used to prepare a cross-sectional lamella for TEM/STEM imaging. TEM was performed using a FEI Titan environmental TEM at 300kV. Classical and 4D-STEM were performed using a Thermo Fisher Spectra monochromated, double corrected STEM at 300kV.

### f. *NV characterization*

All NV center measurements are performed on a home-built confocal microscope using a 532-nm diode laser and an external magnetic field aligned along one of the diamond <111> axes for coherence measurements or misaligned from <111> to reveal all four NV orientations in the scanning CWESR measurements. A cutoff filter of 640 nm is placed in the collection path before the avalanche photodiode to eliminate background fluorescence from $NV^0$. Radiofrequency (RF) signals are delivered through a gold antenna fabricated on a glass slide and placed underneath the diamond. Coherence measurements are performed using a Hahn Echo sequence with a differential measurement scheme.[3] To extract spin densities of NV and P1 centers, XY8 and DEER sequences are used, respectively, and decoherence analysis is applied as in ref. 4, 5.

## II. TEM sample preparation and additional imaging modalities

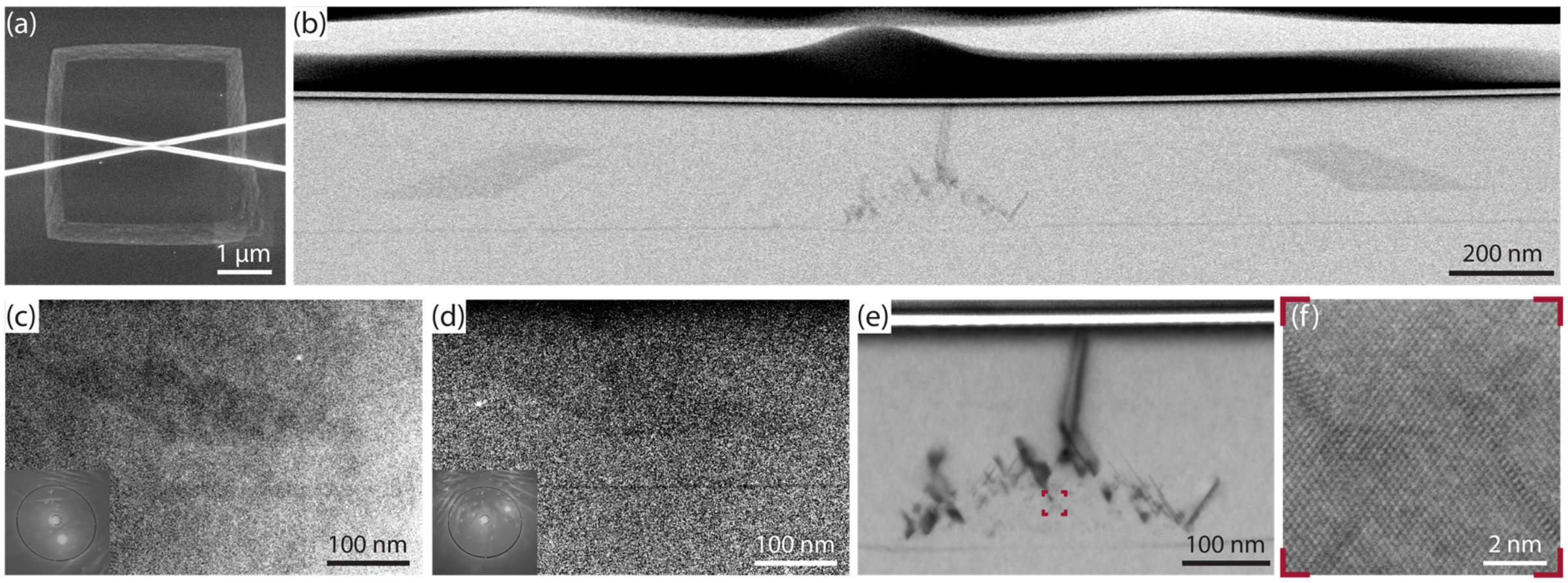

**Fig. S2:** (a) Top-view image of the hillock from which a cross-sectional lamella was prepared. The two Pt lines mark the center of the hillock and were used as references to ensure the center was preserved in the foil. (b) Overview bright field (BF) STEM image of the hillock center at the [110] zone axis (ZA). The dark raised feature in the center top of the image is the center of the Pt cross. TEM BF images of the nitrogen-dense sidewall region to the right of center at the (c) [220] and (d) [004] two-beam conditions. The diffraction patterns for each imaging condition are inset in the bottom left corner. (e) Bright field 4D-STEM reconstruction of the hillock center at [110] ZA, with burgundy corner marks indicating the region of (f) atomic resolution STEM imaging, confirming stacking faults at the center of the hillock.

Fig. S2(a) shows the classic square-shaped hillock from the low miscut sample from which a cross-sectional lamella was prepared for TEM analysis. A Pt cross deposited on the hillock prior to trenching and lift-out was used to mark the center. The center where the Pt overlaps appears as a dark mound at the top center of the foil in (b). The process used to ensure a particular region is preserved for lamella preparation is detailed further in the supplemental information of our prior work.[6] On-zone STEM imaging of the hillock foil reveals the nitrogen-dense sidewall regions both left and right of center (b). The right nitrogen-dense region is also highlighted in the TEM images in (c) and (d) obtained at the g = 220 and g = 004 two-beam conditions, respectively. The visibility of the nitrogen-dense region at orthogonal diffraction conditions indicates that the darkening of this region is due to diffraction contrast associated with a local lattice distortion that we attribute to nitrogen substituting for carbon. The burgundy corner marks in Fig. S2(e) enclose the stacking faults imaged via atomic resolution STEM in (f), providing additional evidence that the angled line features observed at the center of the hillock are stacking faults.

Fig. S3 includes site-coincident AFM 3D renderings (a,e), panchromatic cathodoluminescence (b,f), and $^{15}N$ isotope maps (c,g) of the remaining hillocks from the low-miscut sample analyzed in this work. As with the isolated hillock highlighted in Fig. 3, we observe that the upper and lower sidewalls aggregate more nitrogen than the left and right sidewalls (Fig.S3d,h). Only the leftmost hillock in Fig.S3(b,c) and the rightmost hillock in Fig.S3(f,g) were included in the plot of nitrogen density vs hillock sidewall angle in Fig. 3e as the shorter hillocks in these regions had sides that were engulfed or appeared shadowed by the taller hillock and thus the nitrogen incorporation at these sidewalls was potentially strongly affected by other factors than solely sidewall angle.

## III. Impact of hillock sidewall orientation on nitrogen density

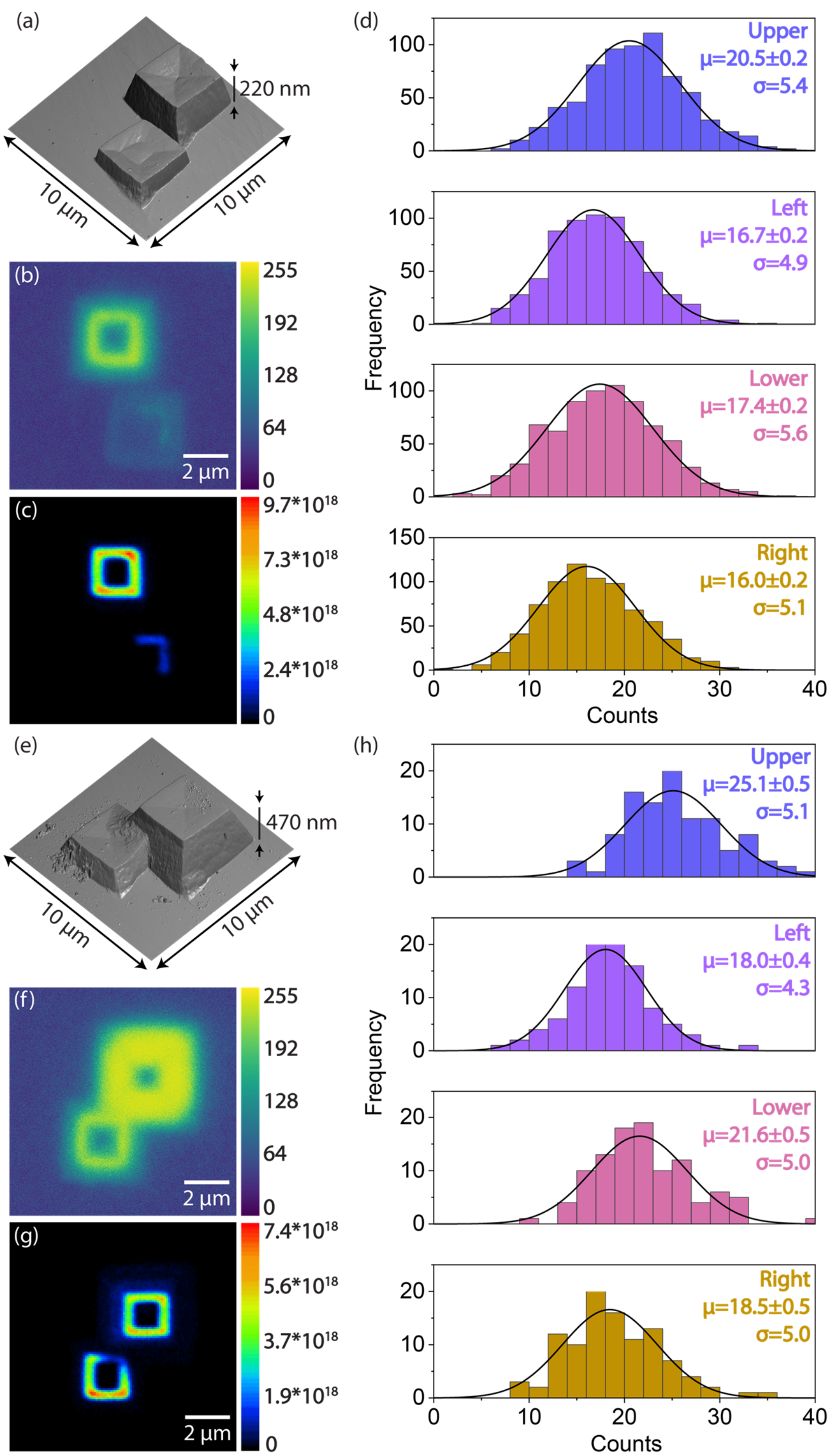

**Fig. S3:** (a,e) AFM 3D-renderings of hillock regions in the low-miscut (θ = 0.60°) sample. (b,f) Panchromatic cathodoluminescence image with colorbar to the right representing grayscale value and (c,g) nanoSIMS $^{15}N$ isotope map with colorbar to the right in units of atoms/cm$^3$ for regions containing two neighboring hillocks (a-c) and two impinging hillocks (e-g). Corresponding histograms (d,h) depict $^{15}N$ counts distribution for the topmost hillock in each region, demonstrating that upper/lower sidewalls have higher $^{15}N$ density compared to left/right sidewalls. Average and standard deviation of a Normal distribution fit are included for each sidewall.

**References**

1. Nuñez, J., Renslow, R., Cliff, J. B. & Anderton, C. R. NanoSIMS for biological applications: Current practices and analyses. *Biointerphases* **13**, (2018).
2. Okazaki, S. Piecewise linear model. (2025).
3. Bluvstein, D., Zhang, Z. & Jayich, A. C. B. Identifying and Mitigating Charge Instabilities in Shallow Diamond Nitrogen-Vacancy Centers. *Phys. Rev. Lett.* **122**, (2019).
4. Davis, E. J. *et al.* Probing many-body dynamics in a two-dimensional dipolar spin ensemble. *Nat. Phys.* **19**, 836–844 (2023).
5. Hughes, L. B. *et al.* Two-dimensional spin systems in PECVD-grown diamond with tunable density and long coherence for enhanced quantum sensing and simulation. *APL Mater.* **11**, (2023).
6. Yan, H. *et al.* Multi-microscopy characterization of threading dislocations in CVD-grown diamond films. *Appl. Phys. Lett.* **124**, (2024).